\documentclass[aps,prl,twocolumn,showpacs,amsmath,amssymb,floatfix,footinbib,superscriptaddress,times]{revtex4-1}
\usepackage{graphics,graphicx,epsfig,ulem,epstopdf,bm,longtable,url,geometry,datetime}
\usepackage[dvipdfm,colorlinks=true,linkcolor=blue,urlcolor=blue,citecolor=blue]{hyperref}
\usepackage{amsmath,amssymb,latexsym,float,amsfonts,subfigure,hyperref,color,physics}
\usepackage[ansinew]{inputenc}
\usepackage[usenames,dvipsnames]{pstricks}
\usepackage[pagewise,mathlines]{lineno}
\def\footnoterule{\kern -1mm \hrule width 6.0cm \kern 2.2mm}%
\usepackage{tikz,xcolor,hyperref}% Make Orcid icon
\definecolor{lime}{HTML}{A6CE39}
\DeclareRobustCommand{\orcidicon}{%
    \begin{tikzpicture}
    \draw[lime, fill=lime] (0,0)
    circle [radius=0.16]
    node[white] {{\fontfamily{qag}\selectfont \tiny ID}};\draw[white, fill=white] (-0.0625,0.095)
    circle [radius=0.007];
    \end{tikzpicture}
    \hspace{-2mm}}
\foreach \x in {A, ..., Z}
{\expandafter\xdef\csname orcid\x\endcsname{\noexpand\href{https://orcid.org/\csname orcidauthor\x\endcsname}{\noexpand\orcidicon}}}

\begin{document}

\title{Multi-timescale time encoding for CNN prediction of Fenna-Matthews-Olson energy-transfer dynamics}
\author{Shun-Cai Zhao\orcidA{}}%
\email[Corresponding author: ]{zsczhao@126.com.}
\affiliation{Center for Quantum Materials and Computational Condensed Matter Physics, Kunming University of Science and Technology, Kunming, 650500, PR China}
\affiliation{School of Science, Department of Physics, Kunming University of Science and Technology, Kunming, 650093, PR China}

\author{Yi-Meng Huang}
\email[Co-first author.]{zscnum1@126.com}
\affiliation{Center for Quantum Materials and Computational Condensed Matter Physics, Kunming University of Science and Technology, Kunming, 650500, PR China}
\affiliation{School of Science, Department of Physics, Kunming University of Science and Technology, Kunming, 650093, PR China}

\author{Yi-Fan Yang}
\affiliation{Center for Quantum Materials and Computational Condensed Matter Physics,  Kunming University of Science and Technology, Kunming, 650500, PR China}
\affiliation{School of Science, Department of Physics, Kunming University of Science and Technology, Kunming, 650093, PR China}

\author{Zi-Ran Zhao}
\affiliation{Center for Quantum Materials and Computational Condensed Matter Physics,  Kunming University of Science and Technology, Kunming, 650500, PR China}
\affiliation{School of Science, Department of Physics, Kunming University of Science and Technology, Kunming, 650093, PR China}
\date{\today}%\currenttime,~  %%\today
\begin{abstract}

%Machine--learning models of open quantum dynamics often rely on recursive predictors that accumulate error. We develop a non--recursive convolutional network that maps system parameters and a redundant time encoding directly to excitation--energy--transfer populations in the Fenna--Matthews--Olson complex. The encoding, based on logistic and $\tanh$ functions, normalizes time and distinguishes fast, transitional, and steady regimes, while physics--informed labels enforce population conservation and inter-site consistency. Trained only on 0\(\sim\)7 \(ps\) trajectories generated with a Lindblad model in QuTiP, the network reliably predicts 0\(\sim\)100 \(ps\) dynamics across reorganization energies, bath rates, and temperatures. Beyond 20 \(ps\), the absolute relative error remains below 0.05, confirming robust long--time extrapolation. By avoiding step--by--step recursion, the approach suppresses error accumulation and transfers across timescales. These results demonstrate that multi-timescale encoding enables data-efficient inference of quantum dissipative dynamics in realistic pigment--protein complexes, providing a route toward machine--learning--assisted design of light-harvesting systems.

Machine learning simulations of open quantum dynamics often rely on recursive predictors that accumulate error. We develop a non-recursive convolutional neural networks (CNNs) that maps system parameters and a redundant time encoding directly to excitation-energy-transfer populations in the Fenna-Matthews-Olson complex. The encoding-modified logistic plus $\tanh$ functions-normalizes time and resolves fast, transitional, and quasi-steady regimes, while physics-informed labels enforce population conservation and inter-site consistency. Trained only on 0\(\sim\)7 \(ps\) reference trajectories generated with a Lindblad model in QuTiP, the network accurately predicts 0\(\sim\)100 \(ps\) dynamics across a range of reorganization energies, bath rates, and temperatures. Beyond 20 \(ps\), the absolute relative error remains below 0.05, demonstrating stable long-time extrapolation. By avoiding step-by-step recursion, the method suppresses error accumulation and generalizes across timescales. These results show that redundant time encoding enables data-efficient inference of long-time quantum dissipative dynamics in realistic pigment-protein complexes, and may aid the data-driven design of light-harvesting materials.

\end{abstract}
%%%%%%%%%%%%%%%%%%%%%%%%%%%%%%%%%%%%%%%%%%%%%%%%%%%%%%%%%%%%%%%%%%%%%%%%%%%%%%%%%%%%%%%%%%%%%%%%%%%%%%%%%%%%%%%%%%%%%%%%%%%%%%%%%%%%%%%%%%%
%%%%%%%%%%%%%%%%%%%%%%%%%%%%%%%%%%%%%%%%%%%%%%%%%%%%%%%%%%%%%%%%%%%%%%%%%%%%%%%%%%%%%%%%%%%%%%%%%%%%%%%%%%%%%%%%%%%%%%%%%%%%%%%%%%%%%%%%%%%
\maketitle
%\tableofcontents
\section{Introduction}\label{Introduction}
%\linenumbers
The nearly 100\% photosynthetic conversion efficiency observed in pigment--protein complexes, such as the Fenna--Matthews--Olson (FMO) complex, has attracted extensive attention \cite{Alharbi2014TheoreticalLO,Li2021InfluenceOT,Li2021Chargetr}. This remarkable efficiency is regarded as a prerequisite for developing artificial photosynthetic devices \cite{Maity2020DFTBMMMD,Schulze2016MultilayerMT,Karafyllidis2017QuantumTI}. Understanding and simulating excitation energy transfer (EET) dynamics in the FMO complex is therefore essential for revealing the underlying physics of its light-harvesting function. EET dynamics are typically described within the framework of open quantum systems, where the reduced density operator encodes system evolution under environmental influence \cite{Li2021Chargetr,Huo2015ElectronicCA,Li2021InfluenceOT}. Several numerically exact methods have been developed to simulate the reduced dynamics, including the hierarchy of equations of motion (HEOM) technique\cite{Tanimura2020}, path-integral Monte Carlo\cite{Kast2012}, multi-configurational time-dependent Hartree (MCTDH)\cite{Meyer1990TheMT}, the stochastic Liouville-von Neumann equation\cite{Stockburger2002}, time-evolving density matrix using orthogonal polynomials algorithm (TEDOPA)\cite{Prior2010}, etc. However, modeling the influence of the environment on quantum systems presents significant challenges due to the vast number of environmental degrees of freedom\cite{RevModPhys.59.1}, limiting their practicality for investigating long-term quantum dynamical phenomena.

In recent years, machine learning (ML) has emerged as a powerful alternative for modeling open quantum dynamics \cite{Hochreiter1997LongSM,Lin2021SimulationOO,Lemm2021MachineLB,Ullah20,Naicker2021MachineLF,Han2021MachineLA,Lin2022AutomaticEO}. Most existing ML approaches adopt recursive strategies \cite{1Ullah2021SpeedingUQ,2Ullah2022OneShotTL,3Wu2021ForecastingND,4Akimov2021ExtendingTT,5Rodrguez2022ACS,6Lin2022AutomaticEO}, where predictions of future states depend on previously predicted values. While conceptually similar to the Markovian propagation of density matrices \cite{Ullah2021SpeedingUQ,Rodrguez2022ACS,20ConvolutionalNN,HerreraRodrguez2024AST}, recursive models are prone to error accumulation, overfitting to short-time data, and numerical instabilities such as vanishing or exploding gradients. These issues can result in violations of basic physical constraints, including trace preservation and positivity of the reduced density matrix. To overcome these limitations, convolutional neural networks (CNNs) \cite{20ConvolutionalNN,Ullah20} have been proposed as non-recursive predictors of quantum dissipative dynamics.

In this work, we extend this idea by developing a refined spatiotemporal mapping approach that integrates QuTiP-based Lindblad simulations of the FMO complex \cite{Johansson2011QuTiPAO} with a redundant time encoding. Specifically, we employ a dual time representation combining modified logistic and $\tanh$ functions to capture short-time relaxation, intermediate transfer, and long-time quasi-steady regimes simultaneously.

In addition, we construct physics-informed labels that embed population conservation and inter-site correlations of the FMO complex, thereby supplying the neural network with robust physical priors.
This design enables accurate predictions of EET dynamics up to 100~\(ps\), even though the model is trained only on short-time trajectories (0\(\sim\)7~\(ps\)).
As shown in this study, the proposed CNN architecture achieves stable long-time extrapolation with absolute relative errors remaining below 0.05 beyond 20~\(ps\), effectively suppressing error accumulation and demonstrating reliable generalization across environmental parameters.

\section{Theoretical Model and Methods}\label{Theoretical Model and methods}
\subsection{Overview and problem statement}\label{sec:overview}

Following the motivation set out in the Introduction, our goal is to predict multi--time--scale excitation--energy--transfer (EET) dynamics in the Fenna--Matthews--Olson (FMO) complex with high fidelity and low computational cost. To bridge the disparate temporal regimes (early coherent beats, mid-time relaxation, and long-time steady behavior), we place a dedicated \emph{redundant time encoding} module at the front of the pipeline, followed by the physical model and the data-generation protocol, and finally the convolutional neural network (CNN) architecture and training strategy. We predict seven site populations $\rho_{jj}(t), j=1,\dots,7$ together with two global summary observables,

\begin{equation}
S_{\mathrm{adj}}(t)=\sum_{i=1}^{6} \lvert \rho_{ii}(t)-\rho_{i+1,i+1}(t)\rvert,~
S_{\mathrm{sum}}(t)=\sum_{i=1}^{7}\rho_{ii}(t),                                        \label{1}
\end{equation}
which enhance sensitivity to inter-site imbalance and provide a population trace check.

\subsection{Redundant time encoding}\label{subsec:time-encoding}

Directly using raw time as a network input is brittle under wide dynamic ranges and uneven resolution demands. We therefore encode $t$ using a bank of smooth, overlapping basis functions that normalize and redundantly represent temporal information. Inspired by Ref.~\cite{Ullah20}, we define for $k=0,1,\dots,99$ and discrete sampling instants $t_n$:
\vspace{-0.3cm}
\begin{align}
f_k(t_n) &= \frac{\eta_k(t_n)+g_k(t_n)}{14},                              \label{eq2}\\
\eta_k(t_n) &= \tanh\!\left(\frac{t_n}{4}+\frac{k}{35}\right),            \label{eq3}\\
g_k(t_n) &= \frac{13}{1+15\exp\bigl[-0.02 (40 t_n + 4k - 1)\bigr]}.       \label{eq4}
\end{align}

Each $f_k(t_n)$ is an S-shaped curve that maps $t$ into $[0,1]$ while providing localized sensitivity. The ensemble of 100 curves acts as overlapping sliding windows, ensuring that each instant is covered by multiple basis functions. Such redundancy enhances robustness: if part of the basis is perturbed by parameter fluctuations or noise, neighboring $f_k$ still encode usable temporal features.

\begin{figure}[htbp]
\centering
\includegraphics[width=0.48\columnwidth]{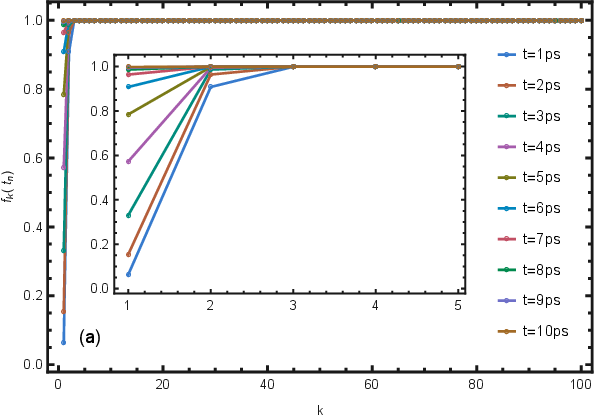 }\includegraphics[width=0.48\columnwidth]{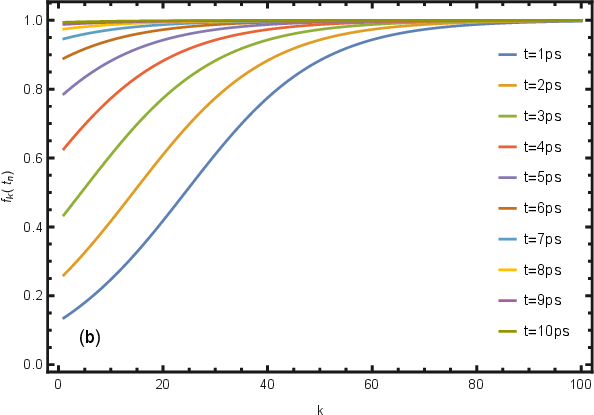 }
\caption{Redundant time-functions $f_{k}(t_{n})$ for $k=0,\dots,99$. (a) Encoding of Ref.~\cite{Ullah20} (inset: $k \in [0,5]$), which achieves normalization only after $k\geq 3$. (b) Present encoding, exhibiting stable normalization and improved discrimination across short-, intermediate-, and long-time regimes.}\label{Fig1}
\end{figure}

Fig.~\ref{Fig1} highlights the difference between the two encodings. In Ref.~\cite{Ullah20}, only part of the basis is properly normalized, leading to reduced sensitivity in early- and long-time regimes [Fig.~\ref{Fig1}(a)]. In contrast, the present design achieves uniform normalization for all $k$ [Fig.~\ref{Fig1}(b)], thereby ensuring consistent resolution from rapid initial transients to late-time steady states. This improvement is crucial for time-dependent quantum dynamics: it avoids overfitting in localized intervals, prevents underrepresentation at large times, and guarantees full coverage of dynamical processes, including relaxation endpoints. Consequently, the refined redundant time-functions serve as stable input features for the CNN (Sec.~\ref{subsec:cnn}), enabling accurate long-horizon predictions of density-matrix evolution and related observables.

\subsection{Model Hamiltonian of the FMO complex}

We model the single-excitation manifold of the FMO complex by the Hamiltonian
\vspace{-0.2cm}
\begin{eqnarray}
\hat{H}_{t}&=&\hat{H}_{e}+\hat{H}_{ph}+\hat{H}_{el-ph},                                                                         \label{eq5}\\
\hat{H}_e&=&\sum\limits_{j=1}^{7} \epsilon_j|j\rangle \langle j| + \sum\limits_{h=1,h \neq j}^{7} J_{jh}|j\rangle \langle h|,   \label{eq6}\\
\hat{H}_{ph}&=&\sum_{\xi}\hbar\omega_{\xi} \hat{b}^{\dagger}_{\xi} \hat{b}_{\xi},                                                \label{eq7}\\
\hat{H}_{el-ph}&=&\sum\limits_{j=1}^{7} (\lambda_{j}+\hat{u}_{j})|j\rangle \langle j|,                                          \label{eq8}\\
{\hat{u}}_{j}&=&-\sum_{\xi}c_{j\xi}\hat{q}_{\xi},                                                                                \label{eq9}
\end{eqnarray}
where the Hamiltonian is separated into electronic, phononic, and electron--phonon interaction parts~\cite{Ishizaki2009TheoreticalEO}.

\begin{figure}[h]
\centering
\includegraphics[width=0.45\columnwidth]{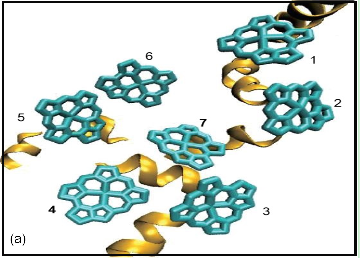}\includegraphics[width=0.53\columnwidth]{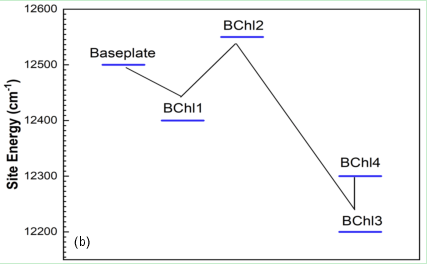}
\caption{(a) Seven bacteriochlorophyll (BChl) pigments in an FMO monomer.
(b) Dominant excitation energy transfer (EET) pathway: baseplate $\rightarrow$ BChl 1 $\rightarrow$ 2 $\rightarrow$ 3 $\rightarrow$ 4, with the initial excitation at BChl 1.}
\label{Fig2}
\end{figure}

Each pigment is modeled as a two-level system: $\ket{j}$ denotes the excited state of BChl $j$, $\epsilon_j$ its site energy, and $J_{jh}$ the electronic coupling to nearby pigments [Eq.~\eqref{eq6}]. These couplings establish the main EET pathways shown in Fig.~\ref{Fig2}(b).

The phonon Hamiltonian $\hat{H}_{ph}$ [Eq.~\eqref{eq7}] describes the protein environment as a set of harmonic modes with frequencies $\omega_\xi$ and bosonic operators $\hat b_\xi$. The interaction Hamiltonian $\hat{H}_{el-ph}$ [Eq.~(\ref{eq8})] incorporates local system--bath coupling, characterized by reorganization energies $\lambda_j$ and coupling constants $c_{j\xi}$ to phonon coordinates $\hat q_\xi$ [Eq.~\eqref{eq9}]. Each site interacts with an independent phonon bath, represented by a Drude--Lorentz spectral density
\begin{equation}
J_j(\omega) = 2 \lambda_j \frac{\omega \gamma_j}{\omega^2 + \gamma_j^2},                                                          \label{eq10}
\end{equation}
where $\gamma_j$ denotes the bath relaxation rate. This spectral density captures both rapid fluctuations and slower environmental relaxation.

To benchmark the model, reference quantum trajectories of the reduced density operator are generated with a local--thermalizing Lindblad master equation~\cite{Mohseni2008}, using the widely adopted Adolphs--Renger seven-site Hamiltonian~\cite{Adolphs2006HowProteins}:
\[
\footnotesize
\begin{bmatrix}
200 & -87.7 & 5.5 & -5.9 & 6.7 & -13.7 & -9.9 \\
-87.7 & 320 & 30.8 & 8.2 & 0.7 & 11.8 & 4.3 \\
5.5 & 30.8 & 0 & -53.5 & -2.2 & -9.6 & 6.0 \\
-5.9 & 8.2 & -53.5 & 110 & -70.7 & -17.0 & -63.3 \\
6.7 & 0.7 & -2.2 & -70.7 & 270 & 81.1 & -1.3 \\
-13.7 & 11.8 & -9.6 & -17.0 & 81.1 & 420 & 39.7 \\
-9.9 & 4.3 & 6.0 & -63.3 & -1.3 & 39.7 & 230
\end{bmatrix}.
\]

All energies are given in cm$^{-1}$. This Hamiltonian captures the essential excitonic structure of the FMO complex and serves as a standard reference for EET dynamics.

\subsection{Dataset construction and feature/label design}\label{subsec:data}

\begin{table}[htbp]
\centering
\vspace{0.3em}
\setlength{\tabcolsep}{0.2cm}\vspace{0.3em}
\resizebox{0.95\columnwidth}{!}{%
\begin{tabular}{|c|c|c|c|c|c|c|c|}
\hline
\textbf{site/10} & $\gamma_a$ & $\lambda_b$ & $T_c$ & $f_k(t_m)$ & $\rho_{ii}(t_m)$ & $\sum_{i=1}^{6}|\rho_{ii}-\rho_{i+1,i+1}|$ & $\sum_{i=1}^{7}\rho_{ii}(t_m)$ \rule{0pt}{3.5ex} \\ % µ÷Õû¸ß¶È
\hline
0.1 & $\gamma_1$ & $\lambda_1$ & $T_1$ & $f_k(t_0)$ & $\rho_{11}(t_0)$ & $|\rho_{11}-\rho_{22}|,\dots$ & $\rho_{11}+\rho_{22}+\dots$ \\
\vdots & \vdots & \vdots & \vdots & \vdots & \vdots & \vdots & \vdots \\
0.1 & $\gamma_1$ & $\lambda_1$ & $T_1$ & $f_k(t_m)$ & $\rho_{11}(t_m)$ & $|\rho_{11}-\rho_{22}|,\dots$ & $\rho_{11}+\rho_{22}+\dots$ \\
\vdots & \vdots & \vdots & \vdots & \vdots & \vdots & \vdots & \vdots \\
\hline
\multicolumn{5}{|c|}{\textbf{Features}} & \multicolumn{3}{c|}{\textbf{Labels}} \\
\hline
\end{tabular}}
\caption{Training schema for the CNN. Time encoding uses Eqs.~\eqref{eq2}--\eqref{eq4} with $k=0,\dots,99$; $t_m = t_0,\dots,t_M$. Environmental grids: $\gamma_a = \gamma_0,\dots,\gamma_J$, $\lambda_b = \lambda_0,\dots,\lambda_J$, and $T_c = T_0,\dots,T_J$.}
\label{Tab}
\end{table}

We assemble supervised training pairs $(\mathbf{x}(t_n), \mathbf{y}(t_n))$ at discrete times $t_n$. The input features $\mathbf{x}(t_n)$ integrate both system-specific physical parameters and temporal encodings, thereby providing the CNN with comprehensive descriptors of excitation energy transfer (EET) dynamics:
\[
\resizebox{.85\columnwidth}{!}{$
\mathbf{x}(t_n) = \bigl[\underbrace{\text{site\_tag}}_{ \in [0.1,\dots,0.7]},\; \underbrace{\lambda/100, \gamma/1000, T/1000}_{\text{ environment}},\; \underbrace{f_0(t_n),\dots,f_{99}(t_n)}_{\text{ redundant time functions}}\bigr].
$}
\]
Here, the \textit{site\_tag} encodes the column index of the reduced density matrix $\rho_{jj}(t)$ as discrete values $\{0.1, \ldots, 0.7\}$, serving as a compact identifier of the excitonic site under consideration. The reorganization energy $\lambda$ characterizes the coupling strength between the chromophoric system and its environment, directly influencing the efficiency of energy transport.
The characteristic frequency $\gamma$ describes the environmental relaxation rate, reflecting the timescale on which the bath responds to electronic excitation. The temperature $T$ accounts for thermal fluctuations and decoherence effects, which critically shape the efficiency and robustness of quantum energy transfer. To capture temporal evolution across multiple scales, we further include $100$ redundant time-functions $f_k(t_n)$ ($k=0,\dots,99$), constructed from logistic basis functions, which normalize the temporal domain and allow the CNN to learn both short-lived quantum beats and long-term asymptotic convergence.

The learning targets $\mathbf{y}(t_n)$ comprise not only the site populations but also two aggregate observables designed to enrich the model¡¯s sensitivity to dynamical features:
\[
\resizebox{.75\columnwidth}{!}{$
\mathbf{y}(t_n) = [\rho_{11}(t_n),\dots,\rho_{77}(t_n), S_{\mathrm{adj}}(t_n), S_{\mathrm{sum}}(t_n)].
$}
\]
Specifically, $S_{\mathrm{adj}}(t_n) = \sum\limits_{i=1}^{6} \bigl|\rho_{ii}(t_n)-\rho_{i+1,i+1}(t_n)\bigr|$ quantifies the fluctuation in exciton populations between adjacent sites, thereby improving the CNN¡¯s ability to detect imbalances and capture energy-transfer transitions.
Meanwhile, $S_{\mathrm{sum}}(t_n)=\sum\limits_{i=1}^{7}\rho_{ii}(t_n)$ monitors the total population within the system, serving as a normalization and fidelity check against probability leakage.
In contrast to Ref.~\cite{Ullah20}, where only the diagonal density matrix elements were used as labels, our expanded output set provides a more comprehensive representation of EET dynamics.

To ensure numerical stability and consistent scaling across features, a systematic preprocessing scheme was employed. All site populations were scaled down by a factor of $10$, yielding normalized labels in the range $0.1$ to $0.7$. The physical parameters were likewise normalized: $\lambda$ was divided by $100$, while $\gamma$ and $T$ were each divided by $1000$. The redundant time-functions $f_k(t_n)$ provide additional normalization in the temporal domain. This preprocessing effectively balances feature magnitudes, stabilizes training, and enhances the CNN¡¯s capacity to accurately learn the underlying physical processes governing EET. The complete schema of features and labels employed in training is summarized in Tab.~\ref{Tab}.

\subsection{CNN architecture and training}\label{subsec:cnn}
\paragraph*{Input layout.} For each site tag and time $t_n$, the input is a 1D vector of length $L=4+100$ (three environment scalars + one site tag + 100 time encoders). Mini-batches stack such vectors along the batch axis.

\paragraph*{Network.} We adopt a one-dimensional CNN to extract temporal and physical features. Its architecture consists of: (i) two convolutional layers (Conv1D $\to$ ReLU), where the first layer maps one input channel to 90 output channels with kernel size 3, and the second layer maps 90 to 70 channels with kernel size 3 and same padding; (ii) a max-pooling layer (MaxPool1d) with kernel size 2; (iii) a flattening layer; and (iv) a multi--layer perceptron (MLP) composed of four fully connected layers (Linear). The first three fully connected layers project the features into a 512-dimensional space and apply ReLU activations, while the final layer maps the representation to the output dimension: ${\left\lfloor \frac{\text{input\_size}-2}{2} \right\rfloor \times70}$. This architecture balances expressive capacity and computational efficiency for predicting the site populations and auxiliary constraints.

For reporting, we use the absolute relative error (ARE),
\[
\mathrm{ARE}(i,t) = \frac{|\rho_{ii}(t)_{\tiny\text{Theo. values}} - \rho_{ii}(t)_{\tiny\text{CNN values}}|}{|\rho_{ii}(t)_{\tiny\text{Theo. values}}|},
\]
and loss populations,
\[
\rho(t)_{\tiny\text{loss}}=\rho_{ii}(t)_{\tiny\text{Theo. values}} - \rho_{ii}(t)_{\tiny\text{CNN values}},~ i \in (1,7)
\]
to revalue the predictive reliability of the proposed CNN architecture.

\paragraph*{Training protocol.} Parameters are optimized using Adam with decoupled weight decay, gradient clipping, and early stopping on a validation split. Inputs are standardized per feature, and labels retain the scaling described in Sec.~\ref{subsec:data}. Model selection is based on the best validation ARE aggregated over all outputs.

\subsection{Implementation details and reproducibility}\label{subsec:impl}
All simulations of reference dynamics and neural training were performed with fixed random seeds and documented preprocessing scripts. Hyperparameter grids (learning rate, batch size, and loss weights) and the exact train/validation/test splits are provided in the (\href{https://github.com/zsczhao/Code-for-CNN}{Codes-for-CNN}) to facilitate replication.

\section{Results and discussions}

\subsection{Training model validation}

To assess the reliability of the CNN model, we performed validation using a dataset spanning 0\(\sim\)7 \(ps\), comprising 301 evenly spaced time points. The excitation-energy-transfer (EET) dynamics for each site of the FMO complex were generated using QuTiP. These trajectories, covering 0\(\sim\)7 \(ps\) (\(m \)= \(7\)), were simulated under varying environmental parameters: reorganization energy \(\lambda \in [14,~28] \ \mathrm{cm^{-1}}\), bath relaxation rate \(\gamma \in [150,~164] \ \mathrm{cm^{-1}}\), and temperature \(T \in [270,~284] \ \mathrm{K}\), each discretized into 15 values. The number of rows in Tab.~\ref{Tab}, the number of feature columns, and the \(j\)-th label column are defined as,

\resizebox{.9\columnwidth}{!}{%
\begin{minipage}{\columnwidth}
\begin{align}
N_\mathrm{rows} &= n_\mathrm{sites} \times n_\mathrm{time} \times n_\gamma \times n_\lambda \times n_T, \label{eq10} \\
N_\mathrm{features} &= n_\mathrm{sites} + n_\gamma + n_\lambda + n_T + n_{f_k(t_m)}, \label{eq11} \\
\mathrm{Labels}_j &= \rho_{jj} + \sum_{i=1}^{6} \left( |\rho_{ii}-\rho_{i+1,i+1}| + (\rho_{ii}+\rho_{i+1,i+1}) \right). \label{eq12}
\end{align}
\end{minipage}%
}

Here, \(n\) denotes the number of corresponding parameters, and \(j\) indicates the \(j\)-th site. Following this procedure, the feature matrix has dimensions \(7{,}111{,}125 \times 104\), and the label matrix has dimensions \(7{,}111{,}125 \times 13\). This matrix representation allows the CNN to efficiently process the dataset while maintaining high predictive accuracy.

The first feature column corresponds to the population of the current energy site, columns 2\(\sim\)7 contain absolute population differences with other sites, and columns 8\(\sim\)13 represent the population sums. The CNN output matrix is compared with the label matrix using absolute relative error (ARE) as the loss function. Iterative back--propagation and optimization yield the trained CNN model .

\begin{figure}[h]
\begin{flushleft}
\includegraphics[width=0.48\columnwidth]{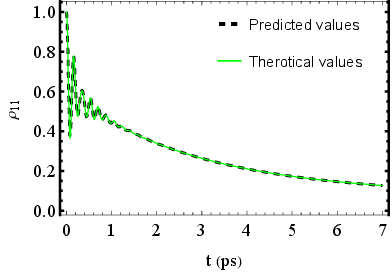}\includegraphics[width=0.48\columnwidth]{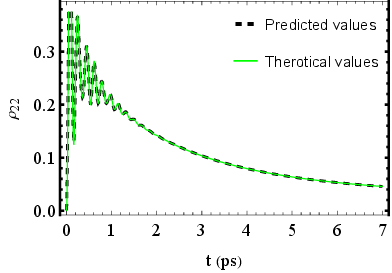}
\includegraphics[width=0.48\columnwidth]{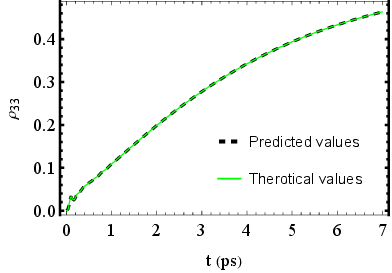}\includegraphics[width=0.48\columnwidth]{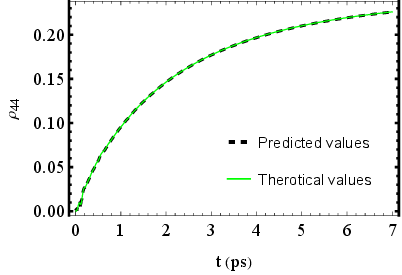}
\includegraphics[width=0.48\columnwidth]{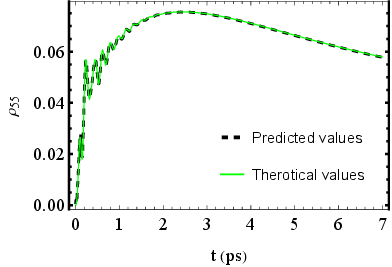}\includegraphics[width=0.48\columnwidth]{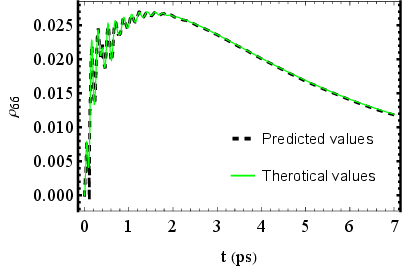}
\includegraphics[width=0.48\columnwidth]{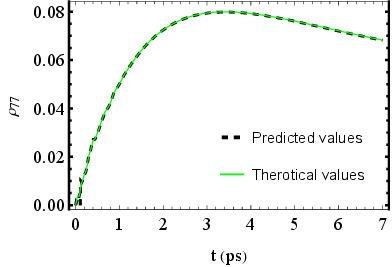}
\end{flushleft}
\vspace{-0.4cm}
\caption{Population dynamics of the seven chlorophyll sites with green solid lines being theoretical values and black dashed lines being predicted values in the FMO complex within 7 \(ps\). Other parameters are $\lambda$=15 $\mathrm{cm^{-1}}$, $\gamma$=275 $\mathrm{cm^{-1}}$, \(T\)=155 $\mathrm{K}$.}\label{Fig3}
\end{figure}

Fig.~\ref{Fig3} demonstrates that the CNN, using refined redundant time-functions, accurately predicts the seven-site evolution over 7  \(ps\). The theoretical and predicted trajectories exhibit near-complete overlap, confirming model fidelity.

Unlike recursive methods~\cite{1Ullah2021SpeedingUQ,HerreraRodrguez2020ConvolutionalNN}, where each step depends on the previous output and is susceptible to error accumulation, our non-recursive CNN predicts EET dynamics directly over the entire temporal domain, reducing computational cost and enhancing long-term stability. The redundant time-functions further enable the model to capture multi-timescale quantum dynamics accurately.

\subsection{Long-term prediction of EET}

Building upon the refined redundant time-functions and distinct label design, we extended the CNN predictions from the training window (0\(\sim\)7 \(ps\)) to a much longer 0\(\sim\)100 \(ps\) timescale. The training trajectories were generated with QuTiP under $\lambda \in [28,30]~\mathrm{cm^{-1}}$, $\gamma \in [164,170]~\mathrm{cm^{-1}}$, and $T \in [284,290]~\mathrm{K}$. This ``short-train, long-predict'' setting directly tests the model's data efficiency and extrapolation capacity.

\begin{figure}[h]
\begin{flushleft}
\includegraphics[width=0.49\columnwidth]{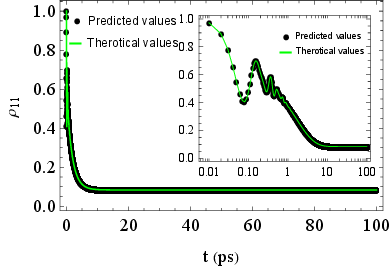}\includegraphics[width=0.49\columnwidth]{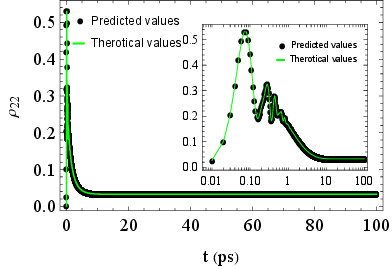}
\includegraphics[width=0.49\columnwidth]{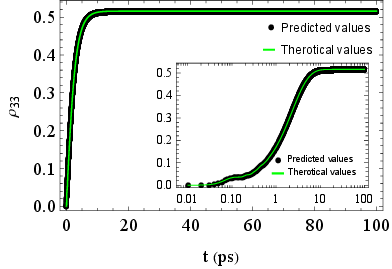}\includegraphics[width=0.49\columnwidth]{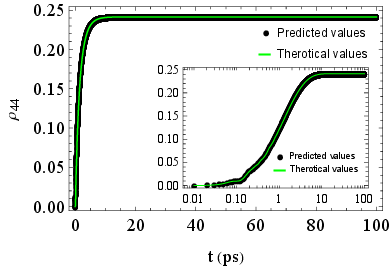}
\includegraphics[width=0.49\columnwidth]{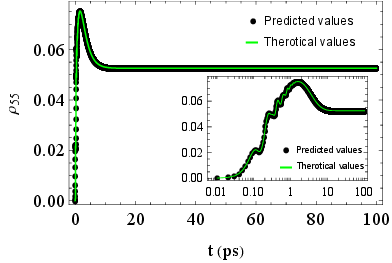}\includegraphics[width=0.49\columnwidth]{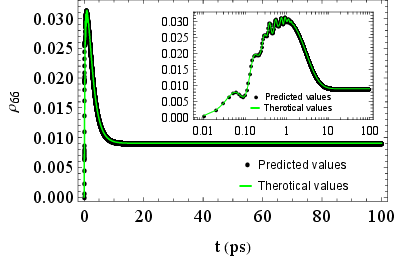}
\includegraphics[width=0.49\columnwidth]{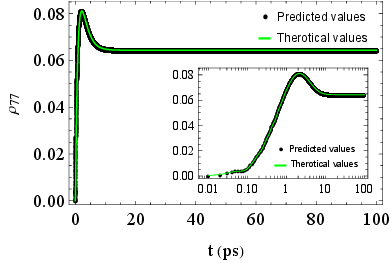}
\end{flushleft}
\caption{Extended population dynamics over 0\(\sim\)100 \(ps\). Green solid lines: theoretical values; black dotted lines: CNN predictions. Insets use logarithmic scales. Parameters: $\lambda=30~\mathrm{cm^{-1}}, \gamma=286~\mathrm{cm^{-1}}, T=166~\mathrm{K}$.}
\label{Fig4}
\end{figure}

Fig.~\ref{Fig4} shows the extended population dynamics for the seven chlorophyll sites. The green solid curves denote theoretical Lindblad results, and the black dotted curves represent CNN predictions. The agreement remains excellent over the entire 0\(\sim\)100 \(ps\) range, with the populations exhibiting early stabilization around 10 \(ps\), consistent with previous reports~\cite{Ullah20}. Insets use logarithmic scales to highlight both rapid transients and long-time saturation. Compared with the 2.5-\(ps\) prediction horizon reported in Ref.~\cite{Ullah20}, our CNN-based approach, empowered by redundant time-encoding, successfully extends the reliable prediction window by nearly two orders of magnitude.

\begin{figure}[h]
\centering
\includegraphics[width=0.75\columnwidth]{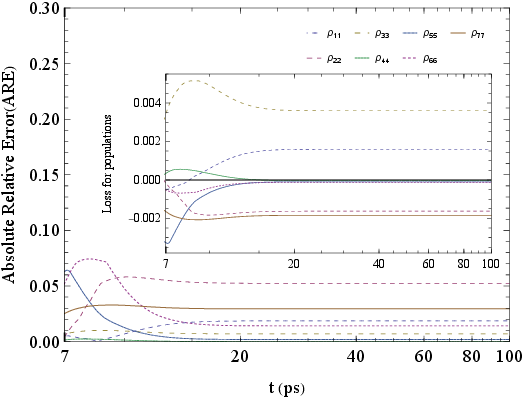}
\caption{Absolute relative error (ARE) and loss function (inset) for seven-site populations over 7\(\sim\)100 \(ps\) (logarithmic scale). Parameters identical to Fig.~\ref{Fig4}.}
\label{Fig5}
\end{figure}

To quantitatively assess stability, Fig.~\ref{Fig5} plots the absolute relative error (ARE) of the seven-site populations for 7\(\sim\)100 \(ps\), with the inset showing the loss function. Although small fluctuations appear in the 7\(\sim\)20 \(ps\) interval, both ARE and loss rapidly settle: beyond 20 ps, the ARE remains below 0.05 while the loss saturates below 0.004. These quantitative indicators demonstrate that our CNN achieves robust long-time extrapolation without recursive accumulation of error.

The performance gain arises from several key design choices. First, convolutional kernels capture local temporal correlations, while translation invariance enables recognition of recurring dynamical patterns across different time windows. Second, hierarchical feature extraction through convolution and pooling reduces dimensionality yet preserves essential dynamical features, improving efficiency and preventing overfitting. Third, the redundant time-encoding normalizes disparate temporal regimes--fast relaxation, intermediate transfer, and quasi-steady evolution--allowing accurate prediction across timescales. Finally, physics-informed labels enforcing population conservation and inter-site consistency stabilize training and enhance fidelity. Together, these factors enable reliable long-term prediction of EET dynamics, outperforming previous recursive and short-horizon models~\cite{Ullah20,1Ullah2021SpeedingUQ,HerreraRodrguez2020ConvolutionalNN}.

\section{CONCLUSION}\label{sec:conclusion}

We presented a non-recursive convolutional framework that combines a redundant, multi-timescale time encoding (modified logistic plus $\tanh$) with physics-informed labels to predict seven-site excitation-energy-transfer dynamics in the Fenna--Matthews--Olson complex. Trained only on short-time reference trajectories (0\(\sim\)7 \(ps\)) generated with a Lindblad model in QuTiP, the network accurately extrapolates the populations to 0\(\sim\)100 \(ps\) across a range of reorganization energies, bath rates, and temperatures. Quantitatively, the absolute relative error remains below 0.05 beyond 20 \(ps\), evidencing stable long-time predictions without step-by-step recursion.

The redundant time encoding normalizes heterogeneous temporal scales---fast relaxation, intermediate transfer, and quasi-steady regimes---so that the CNN can learn time-local features that generalize across the full trajectory. The physics-informed labels, which encode population conservation and inter-site consistency, further regularize training and improve physical fidelity. Together, these design choices suppress error accumulation typical of recursive predictors and deliver data-efficient inference for dissipative quantum dynamics.

Methodologically, the approach is generic: alternative dynamical solvers (e.g., HEOM) can furnish higher-accuracy short-time labels, and the same architecture can be applied to other pigment--protein complexes or to non-Markovian baths where long-time simulation is costly. By enabling robust, quantitatively verifiable long-time predictions from short-time data, our framework offers a practical path toward ML-assisted modeling and design of light-harvesting materials.

%We have introduced a CNN-based framework with novel redundant time-functions and physically informed labels for predicting seven-site EET dynamics in the FMO complex over 100 \(ps\). By eliminating recursive propagation, the model achieves robust long-term predictions and quantitatively reproduces theoretical benchmarks. This approach provides a generalizable architecture for simulating dissipative quantum systems and offers a scalable pathway for optimizing artificial light-harvesting devices under various environmental conditions.

\section*{Code availability}

The code is available at: \href{https://github.com/zsczhao/Code-for-CNN}{Codes-for-CNN} and \href{https://github.com/zsczhao/2-Codes-for-all-Figs}{Codes-for-all-Figs}.

\section*{Author contributions}

S. C. Zhao conceived the idea. Y. M. Huang performed the numerical computations and wrote the draft, and S. C. Zhao did the analysis and revised the paper. Y. F. Yang and Z. R. Zhao participated in the revision and discussion.

\section{Acknowledgment}

We thank the financial supports from the National Natural Science Foundation of China ( Grant Nos. 62065009 and 61565008 ), and the General Program of Yunnan Applied Basic Research Project, China ( Grant No. 2016FB009 ).

\section*{Data Availability Statement}

This manuscript has associated data in a data repository. [Authors' comment: All data included in this manuscript are available upon resonable request by contacting with the corresponding author.]

\section*{Conflict of Interest}

The authors declare that they have no conflict of interest. This article does not contain any studies with human participants or animals performed by any of the authors. Informed consent was obtained from all individual participants included in the study.

\bibliography{reference}
\bibliographystyle{apsrev4-1}%abbrv, acm, unsrt, apalike, ieeetr, apsrev4-1, siam, alpha, plain
\end{document}